\documentclass[
reprint,
amsmath,amssymb,
aps,
]{revtex4-1}
\usepackage[dvipsnames]{xcolor}
\usepackage{graphicx}
\usepackage{dcolumn}
\usepackage{bm}
\usepackage{hyperref}

\begin{document}
	
	
	\title{Bulk-Boundary Correspondence in non-Hermitian Systems:\\ Stability Analysis for Generalized Boundary Conditions }	
	\author{Rebekka Koch$^{1,2}$}
	\email{r.koch@uva.nl}
	\author{Jan Carl Budich$^2$}
	\affiliation{$^1$Institute of Physics, University of Amsterdam, 
		Science Park 904,
		1098 XH Amsterdam, The Netherlands\\
		$^2$Institute of Theoretical Physics, Technische Universit\"{a}t Dresden and W\"urzburg-Dresden Cluster of Excellence ct.qmat, 01062 Dresden, Germany
	}
	
	\date{\today}
	
	\begin{abstract}
			The bulk-boundary correspondence (BBC), i.e. the direct relation between bulk topological invariants defined for infinite periodic systems and the occurrence of protected zero-energy surface states in finite samples, is a ubiquitous and widely observed phenomenon in topological matter. In non-Hermitian generalizations of topological systems, however, this fundamental correspondence has recently been found to be qualitatively altered, largely owing to the sensitivity of non-Hermitian eigenspectra to changing the boundary conditions. In this work, we report on two contributions towards comprehensively explaining this remarkable behavior unique to non-Hermitian systems with theory. First, we analytically solve paradigmatic non-Hermitian topological models for their zero-energy modes in the presence of generalized boundary conditions interpolating between open and periodic boundary conditions, thus explicitly following the breakdown of the conventional BBC. Second, addressing the aforementioned spectral fragility of non-Hermitian matrices, we investigate as to what extent the modified non-Hermitian BBC represents a robust and generically observable phenomenon.  
	\end{abstract}
	
		\maketitle
	
\section{Introduction}
\label{intro}
In a broad variety of physical situations ranging from classical settings to open quantum systems, non-Hermitian  Hamiltonians have proven to be a powerful and conceptually simple tool for effectively describing dissipation. In the classical context, including optical setups \cite{Szameit2011,Regensburger2012,Wiersig2014,Hodaei2017,Feng2017,Sounas2017,Harari2018,bandres2018topological,Zhao2018Topological,Kremer2019,Ozdemir2019}, electric circuits  \cite{circut1,Circuit2Albert2015,Lee2018,Ez2018,helbig2019observation,HoHeScSaetal2019}, and mechanical meta-materials \cite{Nash2015,brandenbourger2019non,ghatak2019observation}, the equations of motion are naturally determined by a non-Hermitian matrix. This situation may in many cases be mapped to an effective (tight-binding) Hamiltonian \cite{ghatak2019observation,Lee2018}  familiar from the quantum-mechanical modeling of electrons in crystalline solids within the independent particle approximation, but with additional non-Hermitian terms. Conversely, as an effective description for dissipative quantum systems, which, at a fundamental level, are governed by Liouvillian dynamics, similar non-Hermitian models can in several scenarios be directly derived  \cite{Bergholtz2019,LindbladSkin2019}.

\begin{figure}

	\resizebox{0.5\textwidth}{!}{%
		\includegraphics{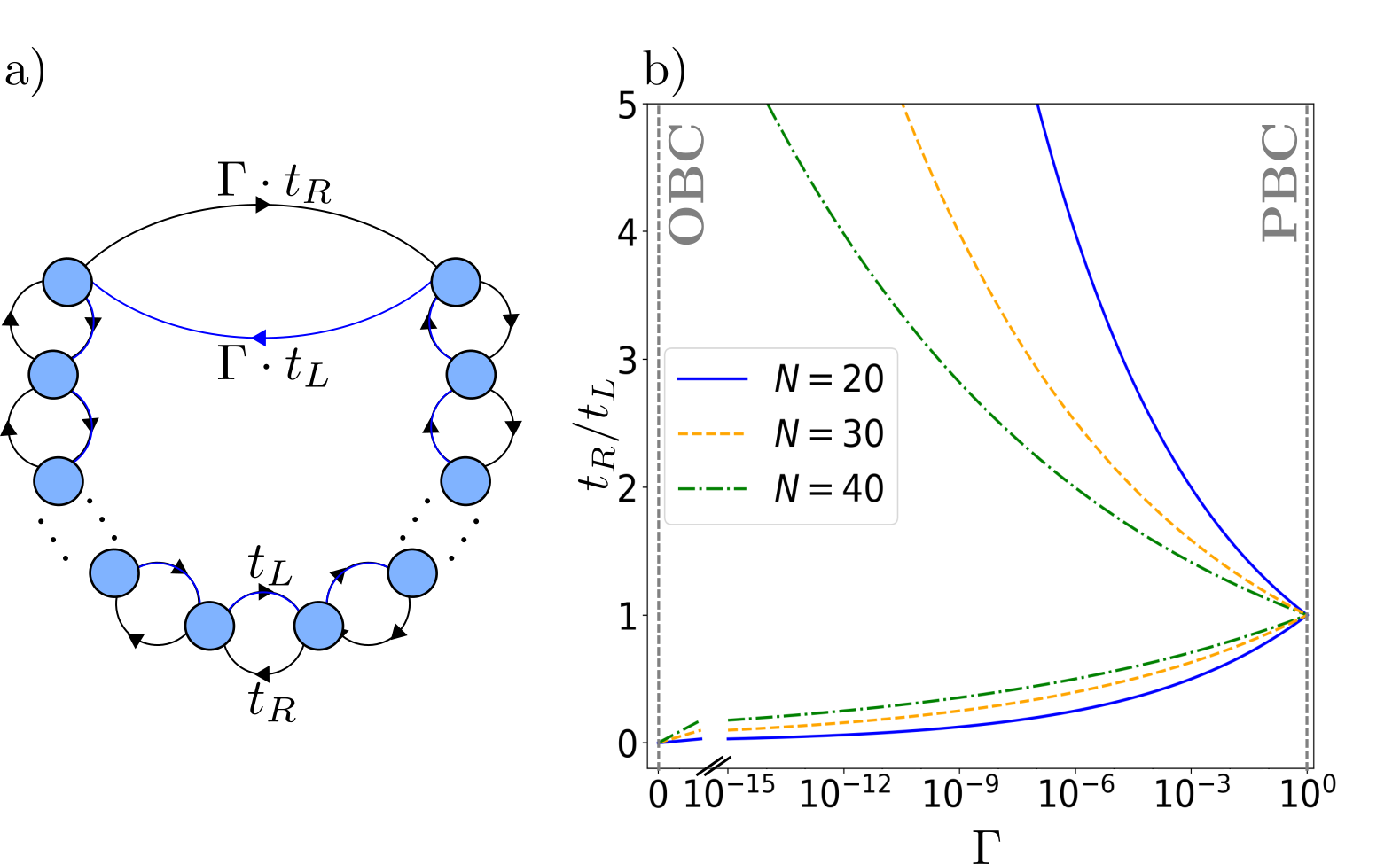}
	}
	
	\caption{The non-Hermitian Hatano-Nelson model is shown as an example for one-dimensional non-reciprocal tight-binding models. In a),  a scheme of the model with generalized boundary conditions that are realized by the parameter $\Gamma \in [0,1]$ is depicted.  In b),  the bulk gap-closing points of the Hatano-Nelson model as a function of the generalized BC parameter ($\Gamma$) for different system sizes are shown.}
	\label{FIGgamma}  
\end{figure}

Recently, a major focus of research has developed on investigating the topological properties of such non-Her\-mi\-tian systems. This pursuit is motivated by both experimental discoveries  \cite{Wang2009,zeuner2015observation,peng2016chiral,weimann2017topologically,chen2017exceptional,zhou2018observation,He2018,xiao2019observation,Cerjan2019,zhou2019exceptional}  and theoretical insights \cite{Esaki2011,gong2018topological,carlstrom2018exceptional,Kawabata2018,Luo2019,budich2019symmetry,yoshida2019,Bergholtz2019,ShenFu,Luitz2019,carlstrom2019knotted,Stalhammar2019,Leykam,edvardsson2019non,ZhouPeriodicTable,torres2019perspective,lee2019hybrid,kawabata2019classification,NewZhang2019correspondence} showing that the notion of topological phases is enriched and quite drastically modified when relinquishing the assumption of hermiticity. In particular, besides the experimental discovery and theoretical prediction of novel non-Hermitian topological phases \cite{gong2018topological,budich2019symmetry,yoshida2019,carlstrom2019knotted,Stalhammar2019}, qualitative changes to the bulk boundary correspondence (BBC) -- a key principle for topological matter -- have been reported \cite{Lee2016halfInt,XiWaWaTo2016,alvarez2018topological,xiong2018does,ghatak2019observation,helbig2019observation,xiao2019observation}: While in Hermitian topological band structures \cite{hasan2010colloquium}, the BBC establishes a one-to-one correspondence between bulk topological invariants characterizing the Bloch bands of an infinite periodic system and the occurrence of protected edge-modes, the breakdown of this direct relation has been demonstrated for various non-Hermitian extensions of topological insulators \cite{Lee2016halfInt,XiWaWaTo2016,xiong2018does,kunst2018biorthogonal}. Motivated by this observation, several approaches to formulate a modified non-Hermitian BBC have been put forward \cite{gong2018topological,kunst2018biorthogonal,yao2018edge,Rosenow2019,herviou2019restoring,jin2019bulk,FeiWang,lee2019anatomy,yokomizo2019bloch,edvardsson2019non,Imura}. Since the mentioned breakdown of the standard BBC is closely related to known instabilities of the eigenvalue spectrum of non-Hermitian matrices \cite{reichel1992eigenvalues}, the physical robustness of the non-Hermitian BBC has been questioned \cite{gong2018topological,kunst2018biorthogonal}, and analyzing the more stable singular value spectrum has been proposed as an alternative diagnostic tool \cite{herviou2019restoring}.\\
In this work, we address several remaining issues regarding the BBC in non-Hermitian systems, focusing on the biorthogonal basis approach reported in Ref. \cite{kunst2018biorthogonal}. First, we demonstrate the robustness of non-Hermitian zero-energy boundary modes against physically relevant local perturbations, despite the aforementioned fragility of the eigenstate spectrum towards adding generic (random) matrices to the non-Hermitian Hamiltonian. Specifically, the matrix-elements of a non-Hermitian tight-binding model are found to typically decay fast enough in real space to overrule the rapidly growing relevance with spatial distance of (non-local) perturbations.  Second, considering one-dimensional (1D) non-Hermitian systems, we derive analytical expressions for the occurrence of exceptional point (EP) transitions and the formation of zero-energy edge modes as a function of a generalized boundary condition parameter $\Gamma$ continuously interpolating between periodic ($\Gamma = 1$) and open ($\Gamma=0$) boundaries (see Fig. \ref{FIGgamma} for an illustration). In this context, the discrepancy between the topological phase diagrams of systems with different boundary conditions is intuitively explained by the occurrence of topological phase transitions, in which the boundary condition parameter $\Gamma$ plays the role of a control parameter. Our findings provide additional insights on both the analytical origin and experimental relevance of anomalous (from a Hermitian perspective) bulk boundary effects in topological non-Hermitian systems.
\\
The remainder of this article is organized as follows: We start by formally introducing generalized boundary conditions in topological non-Hermitian one- and two-band models in Section \ref{Chap_models}. We shortly discuss previous approaches to restore the BBC for non-Hermitian systems, thereby focusing on the biorthogonal framework proposed in Ref.~\cite{kunst2018biorthogonal} in Section \ref{Chap_biorth}. After deriving the analytical  EP transitions for two paradigmatic non-Hermitian models in Section \ref{Chap_Analytics}, the stability of  their eigenspectra and edge modes is discussed  in Section \ref{Chap_stability}. Finally, a concluding discussion is presented in Section \ref{discussion}.

\section{Non-Hermitian tight-binding models with generalized boundary conditions}\label{Chap_models}
We focus on one-dimensional tight-binding models with $N$ unit cells and generally asymmetric hopping amplitudes between the sites that render the Hamiltonian non-Hermitian. Compared to their Hermitian counterparts, some non-Hermitian tight-binding models show qualitatively different eigenspectra depending on the imposed boundary conditions \cite{Lee2016halfInt,XiWaWaTo2016,kunst2018biorthogonal}, i.e. periodic boundary conditions (PBC) or open boundary conditions (OBC). In fact, the whole eigenspectrum and with it all the eigenstates are affected by the boundary conditions: In the OBC case, not only zero-energy edge modes familiar from Hermitian topological systems but also a macroscopic number of bulk modes may be exponentially localized at one of the edges - a phenomenon coined the non-Hermitian skin-effect \cite{YaSoWa2018,yao2018edge}, whose connection to the failure of the conventional BBC has been widely discussed \cite{alvarez2018topological,kunst2018biorthogonal,Kunst2019,Kawabata2018,FeiWang,HoHeScSaetal2019,NewZhang2019correspondence}. Indeed, the non-Hermitian skin effect is found to occur alongside the discrepancy between the PBC and OBC eigenspectrum in these models \cite{jin2019bulk}.\\
To examine the extreme sensitivity towards the boundary conditions, one may introduce generalized boundary conditions by scaling the hopping between  the last site $n=N$ and the first site $n=1$  with a parameter $\Gamma \in [0,1]$ such that  $\Gamma =0$ ($\Gamma=1$) corresponds to OBC (PBC) \cite{xiong2018does,kunst2018biorthogonal} as depicted in Figure \ref{FIGgamma}. \textcolor{black}{Note, that for $\Gamma \ne 0, 1$ this generalized BC construction can also be viewed as an impurity localized between site $1$ and $N$.} It  has been observed \cite{xiong2018does,kunst2018biorthogonal,lee2019anatomy}  (e.g. by using a complex flux to tune the boundary conditions \cite{lee2019anatomy}) that the transition between the periodic and the open system happens almost instantaneously. That is, the eigenspectrum exhibits a qualitative change of order one as soon as  the boundary conditions is modified by the critical parameter $\Gamma_c$ that is exponentially small in system size. In other words, $\Gamma_c \propto \exp(-\alpha N)$ with $\alpha$ some positive real-valued number depending on the  system's parameters is enough to trigger a bulk transition in the spectrum \cite{kunst2018biorthogonal}. 
\\
A minimal model with asymmetric hopping amplitudes  is the Hatano-Nelson model  \cite{Hatano1996}. The real space Hamiltonian reads as \cite{gong2018topological}
\begin{align}
H_{\text{HN}}&=\sum_{j=1}^{N-1}(t_R c^{\dagger}_{j+1}c_j+t_Lc_j^{\dagger}c_{j+1}) +\Gamma (t_R c^{\dagger}_{1}c_N+t_Lc_N^{\dagger}c_{1}).	\label{HNhamiltonian}
\end{align}
where $N $ is the number of sites, $c_j$ ($c_j^{\dagger}$) are the fermionic annihilation (creation) operators on site $j$,  $t_L, t_R\in \mathbb{C}$ are the hopping amplitudes, and $\Gamma \in [0,1]$ determines the boundary conditions. Its Hamiltonian in reciprocal space with PBC ($\Gamma=1$) is given by
\begin{align}
\sum_k c_k^{\dagger}\mathcal{H}(k)c_k \quad \text{with } \mathcal{H}(k)=t_R e^{-ik}+t_Le^{ik}
\end{align} 
where the lattice momentum $k$ is summed over the first Brioullin zone (BZ). While in the Hermitian context, the meaning of a bulk gap is lost for one-band models, the {\textit{complex}} spectrum of the Hatano-Nelson model exhibits a point-gap \cite{gong2018topological}, as long as there is no state at $E=0$, i.e. $\lvert t_R \rvert \ne \lvert t_L \rvert$. This one-band model possesses a point-symmetric eigenspectrum in the complex plane, if the total number of sites is even. Hence,  the eigenenergies come in pairs $\{E, -E\}$ such that the eigenenergy $E=0$ is degenerate for an even number of sites. Moreover, $\mathcal{H}(k)=0$ determines the zero energy, i.e. point-gap closing points \cite{gong2018topological}, for the periodic system in exact analogy to Hermitian systems. \\
Another emblematic class of models is provided by non-interacting two-band models which are fully described by their  Bloch Hamiltonian $\mathcal{H}(k)$ in reciprocal space. The  non-Hermitian version of $\mathcal{H}(k)$  reads as
\begin{eqnarray}
\mathcal{H}(k)&=&\mathbf{d}_{Re}(k)\cdot \textbf{$\sigma$}+ i\mathbf{d}_{Im}(k)\cdot \mathbf{\sigma}, \label{BlochH}
\end{eqnarray}
where  \textbf{$\sigma$}$=(\sigma_x, \sigma_y, \sigma_z)$ are the standard Pauli matrices, $k$ is the lattice momentum and $\mathbf{d}_{Re}$, $\mathbf{d}_{Im} \in \mathbb{R}^3$. Its eigenvalues are given by 
\begin{eqnarray}
E_{\pm}&=& \pm \sqrt{\textbf{d}_{Re}^2-\textbf{d}_{Im}^2+2i\textbf{d}_{Re}\cdot\textbf{d}_{Im}}.\label{eigenvalue}
\end{eqnarray}
Since the Hamiltonian is non-Hermitian, we encounter  points in parameter  space that not only feature degenerate eigenvalues but also their corresponding eigenvectors coalesce. Such points are called exceptional points (EPs) \cite{Heiss2004} that render the Hamiltonian defective (non-diagonalizable). In the case of generic non-Hermitian Bloch-Hamiltonians (see equation (\ref{BlochH}) and (\ref{eigenvalue})), both eigenvalues coincide if $E_+=E_-=0$ which generally is an EP of order two  apart from the trivial case $ \mathbf{d}_{Re}=\mathbf{d}_{Im}=0$ known as the diabolic point. Note, that the one-band equivalent $\mathcal{H}(k)$ for the Hatano-Nelson model has no EPs because a scalar cannot be defective. There, however, the non-Hermitian real-space tight-binding Hamiltonian can still exhibit EPs \textcolor{black}{if $\Gamma\ne 1$}.
\\
The unit-cell of the considered two-band tight-binding models consists of two alternating sites $A, B$. 
Having imposed periodic boundary conditions (PBC), the position space Hamiltonian $H$ is obtained upon Fourier transform with the fermionic creation (annihilation) operators $c^{\dagger}_{k, A(B)}$ ($c_{k, A(B)}$) in $\sum_k (c^{\dagger}_{k, A}, c^{\dagger}_{k, B}) \mathcal{H}(k)(c_{k, A}, c_{k, B})^T$, where the momentum $k$ is summed over the BZ. Thus,  $H$ still encounters an EP at $E=0$ whenever  $\mathcal{H}(k)$ does, which at the same time marks the bulk gap-closing point. \\

\begin{figure}
	
	\resizebox{0.5\textwidth}{!}{%
		\includegraphics{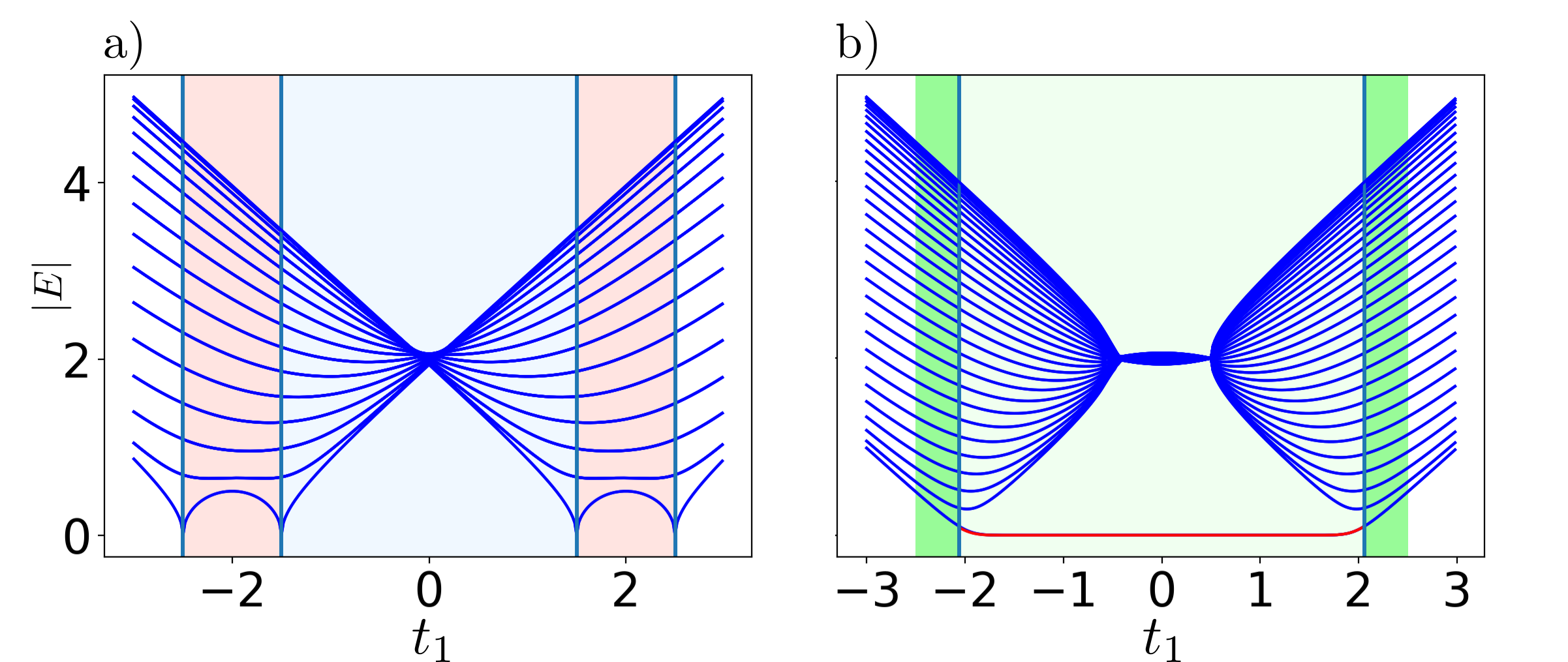}
	}

	\caption{The absolute eigenvalues (blue lines) of the non-Hermitian SSH-model with PBC in a)  and OBC in b)  are shown. The parameters are $t_2=2, \gamma =1$ and the number of unit cells is $N=20$. The red shaded areas show where the eigenenergies encircles the origin $E=0$ and thus are topologically non trivial according to \cite{gong2018topological}. The blue shaded area shows where a half-integer winding number (complex extension of the Chern-number \cite{Lee2016halfInt}) indicates a non-trivial topological phase. The lighter green area shows where the biorthogonal polarization $P=1$ \cite{kunst2018biorthogonal} and edge states depicted in red occur while the darker green shaded areas show where no edge state forms even though the periodic spectrum suggest a topological non-trivial phase (compare to shading in a)). } 
	\label{FigSSHmodel}  
	\end{figure}

We illustrate our further analysis based on a non-Her\-mi\-tian version of the  Su-Schrieffer-Heeger model (NH SSH) \cite{SSH} that has been widely discussed in Refs.~\cite{yao2018edge,kunst2018biorthogonal,lee2019anatomy,herviou2019restoring,jin2019bulk,Imura}, where
\begin{align}
d_x=t_1 +t_2 \cos(k), \quad d_y=i\frac{\gamma}{2}+t_2\sin(k),\quad d_z=0
\end{align}
are the components of $ \textbf{d}$ in equation (\ref{BlochH}).
Since this Hamiltonian preserves the chiral symmetry $\sigma_z\mathcal{H}(k) \sigma_z=-\mathcal{H}(k)$,   the eigenenergies come in pairs $\{E, -E\}$ such that the eigenenergy $E=0$ is degenerate, if the total number of sites is even. In Fig. \ref{FigSSHmodel} the absolute values of the eigenenergies are depicted for $a)$ PBC and $b)$ OBC to show the qualitative differences of both spectra for this model.
\\
The  bulk gap-closing points as a function of the hopping amplitude $t_1$ are determined  by the roots of (\ref{eigenvalue})  in the periodic system
\begin{eqnarray}
t_1=\pm\left(t_2+\frac{\gamma}{2}\right), \quad t_1=\pm\left(t_2-\frac{\gamma}{2}\right),
\end{eqnarray}
while it has been found \cite{kunst2018biorthogonal,yao2018edge} that
\begin{align}
t_1=\pm\sqrt{\frac{\gamma^2}{4}-t_2^2} \text{~~if~}|t_2|<\frac{\gamma}{2}, \quad t_1=\pm\sqrt{t_2^2+\frac{\gamma^2}{4}} \label{SSHobc}
\end{align}
give the bulk gap-closing points for the open system. The NH SSH-model has thus either two ( $t_2>\gamma/2$) or four bulk gap-closing points for OBC \cite{yao2018edge}. Alongside the whole eigenspectrum, the position of the bulk gap-closing points in parameter space is altered as well  when imposing  OBC. The resulting qualitatively differing eigenspectra  of the periodic and the open system (see differences in a) and b) of Figure \ref{FigSSHmodel} as an example) cause  the conventional bulk-boundary correspondence (BBC) to break down \cite{Lee2016halfInt,yao2018edge,kunst2018biorthogonal}.  Usually, this celebrated correspondence connects the occurrence and number of protected surface states of a Hermitian Hamiltonian with open boundaries  to a topological invariant calculated from the bulk Hamiltonian, i.e. the system with PBC. Then, the bulk-band touchings mark the borders between different topological phases \cite{hasan2010colloquium}. \\
Introducing again the generalized boundary conditions parameterized by $\Gamma\in [0, 1]$, the real space NH SSH Hamiltonian explicitly reads as
\begin{align}
&\nonumber H_{SSH}=\left( \sum\limits_{n=1}^{N-1}\left(t_2c^{\dagger}_{n+1, A}c_{n, B}\right)+\Gamma t_2c^{\dagger}_{N, A}c_{1, B}\right)+h.c.\\&+
\sum\limits_{n=1}^N\left(\left(t_1+\frac{\gamma}{2} \right)c^{\dagger}_{n, A}c_{n, B}+\left(t_1-\frac{\gamma}{2} \right)c^{\dagger}_{n, B}c_{n, A}\right).
\label{SSHGamma}	
\end{align}

In Section \ref{Chap_Analytics}, analytical solutions for the bulk transition-points as a function of $\Gamma$ for both the Hatano-Nelson model (\ref{HNhamiltonian}) and the NH SSH model (\ref{SSHGamma}) will be presented.

\section{Biorthogonal bulk boundary correspondence approach}\label{Chap_biorth}
Recently, several suggestions for a non-Hermitian version of the bulk-boundary correspondence have been proposed \cite{gong2018topological,yao2018edge,Rosenow2019,herviou2019restoring,jin2019bulk,FeiWang,lee2019anatomy,yokomizo2019bloch,Imura}, including an approach based on the  biorthogonal polarization $P$  \cite{kunst2018biorthogonal} on which we elaborate below. Accounting for the aforementioned sensitivity of non-Her\-mi\-tian systems to boundary conditions, the crucial difference to conventional bulk topological invariants is that $P$ is calculated from bulk states of a system with OBC, thus accurately predicting bulk-band touching points  and surface states for the open system (light green shaded area in Figure \ref{FigSSHmodel}b)). Similar results have been obtained in a complementary way by  considering a generalized BZ for non-Hermitian systems \cite{yao2018edge} that contains model-specific information on both the periodic and the open boundary system.\\
In contrast, there have been some notable  suggestions for non-Hermitian topological invariants that are deduced solely  from  the periodic spectrum \cite{Esaki2011,Lee2016halfInt,gong2018topological,ShenFu}. Firstly, the non-Hermitian half-integer  winding number \cite{Lee2016halfInt} (blue shaded area in Figure \ref{FigSSHmodel}a)). Secondly, one can construct a different type of winding number based on the   encircling of   complex eigenenergies around the origin $E=0$ in the complex plane that is trivially zero in all Hermitian systems \cite{gong2018topological} (red shaded area in Figure \ref{FigSSHmodel}a)). 
However, both in general do not correctly predict the occurrence and disappearance of edge states in the open system for some parameter regime (compare to Figure \ref{FigSSHmodel} b)).  
\\

\textit{Biorthogonal Polarizaton.}  We now briefly review the construction of the  biorthogonal polarization $P$ \cite{kunst2018biorthogonal} for later reference.
If the Hamiltonian is non-Hermitian,  one has to distinguish between right and left eigenstates $|\psi_{R, i}\rangle$ and $|\psi_{L, i}\rangle$ found from the right and left eigenvalue problem, respectively:
\begin{align}
&H|\psi^R_{ i}\rangle=E_i|\psi^R_{ i}\rangle \text{, and } H^{\dagger}|\psi^L_{ i}\rangle=E^{\ast}_i|\psi^L_{ i}\rangle.
\end{align}
Choosing the biorthogonal normalization $\langle \psi^L_{ i} |\psi^R_{ j}\rangle=\delta_{ij}$ \cite{Brody2013} gives the set $\{ |\psi^R_{ 1}\rangle, |\psi^R_{2}\rangle, \dots, |\psi^L_{ 1}\rangle, |\psi^L_{ 2}\rangle, \dots\}$ that spans the complete eigenspace unless the system is at an EP.
\\
One-dimensional Bloch-Hamiltonians, where the vector $\textbf{d}$ in (\ref{BlochH}) is given by $d_x(k) \pm i d_y(k)= f_{\pm} + g_{\pm} \exp(\pm ik a)$ with $a$ the lattice vector between neighboring unit cells, always possess  an exact eigenmode with energy $E=d_z$,  if the total number of sites  is odd (e.g. the last unit cell is broken) and OBC are imposed \cite{KuMiBe2018w1}.  Assuming the chain starts and ends with an $A$-site, the corresponding exact left and right eigenstates read
\begin{align}
&|\psi_{R}\rangle= \mathcal{N}_R \sum\limits_{j=1}^Nr_R^jc^{\dagger}_{j, A}|0 \rangle, \quad {\color{black}r_R=-\frac{f_+}{g_+}} \label{eigenmodeR} \\
&|\psi_{L}\rangle= \mathcal{N}_L \sum\limits_{j=1}^Nr_L^jc^{\dagger}_{j, A}|0 \rangle, \quad r_L=-\frac{f_-^{\ast}}{g_-^{\ast}} \label{eigenmodeL}
\end{align}
with $N$ the number of unit cells and  $c^{\dagger}_{n, A(B)}$ $[c_{n, A(B)}]$  the creation [annihilation] operators of a particle on sublattice $A(B)$ in unit cell $n$, and where $\mathcal{N}_{R(L)}$ is  a normalization factor according to the biorthogonalization condition. These states are exponentially localized to one of the edges of the chain.  In Ref. \cite{kunst2018biorthogonal}, it has been found that  a topological phase transition in the open system occurs if the biorthogonal analogue of the mode localization  changes from one side to the other. That is, the biorthogonal expectation value of the projector $\Pi_j=c^{\dagger}_{j, A}|0 \rangle  \langle 0|c_{j, A}+c^{\dagger}_{j, B}|0 \rangle  \langle 0|c_{j, B}$ onto the $j$th unit cell 
\begin{align}
&\langle\psi_L|\Pi_j|\psi_R\rangle =\mathcal{N}^{\ast}_L\mathcal{N}_R(r_L^{\ast}r_R)^j.
\end{align} 
is exponentially localized to the left (right) edge for 
$|r_L^{\ast}r_R|>1$ ($|r_L^{\ast}r_R|<1$). Furthermore,  the bulk gap of the open system closes if
\begin{align}
|r_L^{\ast}r_R|=1 \label{floreGapC}.
\end{align}
Using this, one constructs the biorthogonal polarization
\begin{align}
&P=1-\lim\limits_{N\rightarrow \infty}\frac{1}{N}\langle\psi_L|\sum_j j\Pi_j|\psi_R\rangle,
\label{eqn:biopol}
\end{align}
which is quantized in the thermodynamic limit and moreover jumps precisely at the bulk band touching points between $0$ and $1$.  $P$ thus  plays the role of a bulk invariant characterizing systems with OBC by  predicting edge states in the following sense:\\

\noindent\begin{tabular}{lll}
	\hline\noalign{\smallskip}
	& odd $\#$ of sites & even $\#$ of sites\\ 			
	\noalign{\smallskip}\hline\hline\noalign{\smallskip}
	$P=0$	&  left localized edge state& no edge states  \\ 
	\hline 	\rule[0.5ex]{0pt}{2.5ex}
	$P=1$	& right localized edge state  &  two edges states\\ 
	\noalign{\smallskip}\hline
\end{tabular}
\section{Analytical Solution with Generalized BC}\label{Chap_Analytics}
Going beyond previous literature, we now analytically  derive the bulk transition points at the ``critical" generalized boundary conditions $\Gamma_c=\exp(-\alpha N)$ \cite{kunst2018biorthogonal} for the non-Hermitian models introduced in Section \ref{Chap_models}, thereby specifying the dependence of $\alpha$ on model-parameters. To this end, we solve the right and left eigenvalue problems of the position space Hamiltonian $H$ for the specific eigenvalue $E=0$ with generalized boundary conditions. If PBC are imposed ($\Gamma=1$), this ansatz gives the eigenstates associated to the gap-closing points of the Bloch (or Bloch-like) Hamiltonian. By following these eigenstates while $\Gamma$ is continuously interpolated between 1 and 0, the exact parameter dependence of $\Gamma_c$ is revealed. 
\\
We furthermore find an EP at zero energy that moves through parameter space as a function of the boundary condition parameter $\Gamma$ for an even number of sites. In particular, this EP coincides with the gap-closing point whenever there is no edge mode in the OBC spectrum.

\paragraph{Hatano-Nelson model.}

\begin{figure}
	
	\centering

	\resizebox{0.51\textwidth}{!}{%
		\includegraphics{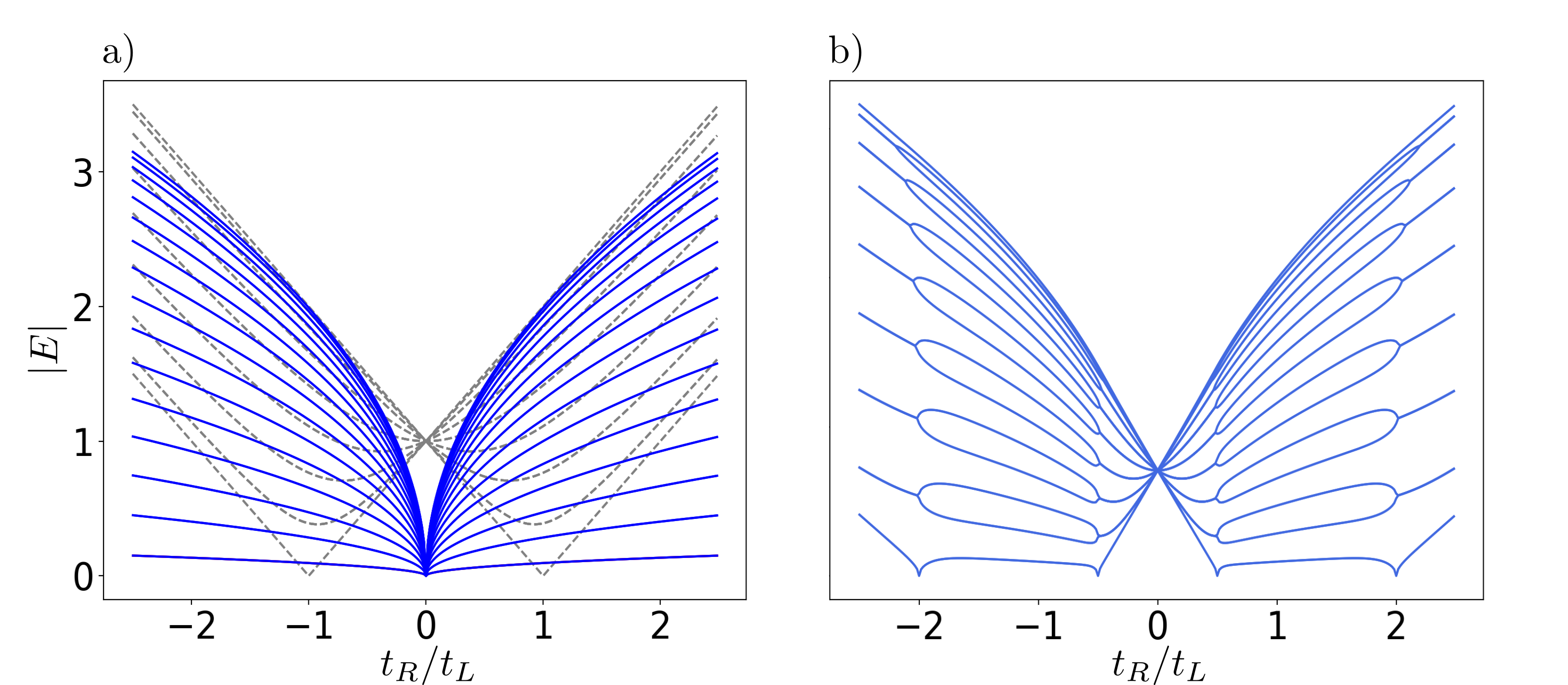}
	}
	
	\caption{The absolute eigenenergies of the Hatano-Nelson model  with $N=32$ sites in dependency of the ratio $t_R/t_L$ are depicted. In a), the difference between imposing open (blue) and periodic (dashed gray) boundary conditions is shown. In b), the ends of the chain are coupled with $\Gamma_c =\exp(-N\ln\sqrt{2})$ such that the bulk gap closes for $t_R/t_L=\pm2$ and $t_R/t_L=\pm1/2$.}\label{FigNelsonHat}
\end{figure}

Solving the right eigenvalue problem $H_{HN}|\psi^R_0\rangle=E|\psi^R_0\rangle$ (see eq. (\ref{SSHobc})) with the ansatz $|\psi^R_0\rangle=\sum_{n=1}^N\psi^R_{n}c^{\dagger}_{n}|0\rangle$  and assuming that the eigenenergy is given by $E=0$, we arrive at the  following  set of equations
\begin{align}
&0=t_L \psi_{n-1}+t_R\psi_{n+1}, \quad\text{for~} n=2, 3, \dots, N-1, \label{sysOfEqHN}
\end{align}
with the boundary conditions 
\begin{align}
&0=t_L\psi_{N-1}+\Gamma t_R\psi_{1}, \quad 0=\Gamma t_L\psi_N+t_R\psi_{2}.\label{sysOfEqHNboundary}
\end{align}
Thus, the Hatano-Nelson model also possesses one exact zero-energy eigenstate throughout the whole parameter range with vanishing amplitudes on all even numbered sites, if the total   number of sites $N$ is odd and OBC ($\Gamma=0$) are imposed. \\
If the number of sites $N$ is even,  we solve the system of equations for $\Gamma $ and we find \textcolor{black}{the two independent solutions labeled with  $a,b$}
\begin{align}
\color{black}
&\Gamma^a_c=\left(-\frac{t_R}{t_L}\right)^{N/2}={\color{black}e^{N\ln\left(\sqrt{\frac{t_R}{t_L}}\right)}=e^{-\frac{N}{\xi_a}}} \in [0,1]\label{gamma1nh}\\
&\Gamma^b_c=\left(-\frac{t_L}{t_R}\right)^{N/2}={\color{black}e^{N\ln\left(\sqrt{\frac{t_L}{t_R}}\right)}=e^{-\frac{N}{\xi_b}}} \in [0,1],
\label{gamma2nh}
\end{align}
\textcolor{black}{since the system of equations (\ref{sysOfEqHN}), (\ref{sysOfEqHNboundary}) decouples with respect to even- and odd-numbered sites.  $\xi_{a(b)}$ are the localization lengths of the corresponding right eigenvectors}  
\begin{align}
&|\psi^R_{\color{black}0,a}\rangle=\mathcal{N}_R\sum_{n=1}^{N/2}\left(-\frac{t_R}{t_L}\right)^{n-1}c_{2n-1}^{\dagger}|0\rangle \label{firstev}\\
&|\psi^R_{\color{black}0,b}\rangle=\mathcal{N}_R\sum_{n=1}^{N/2}\left(-\frac{t_L}{t_R}\right)^{N/2-n}c_{2n}^{\dagger}|0\rangle, \label{secondev}
\end{align}
\textcolor{black}{where $|\psi^R_{0, a(b)}\rangle$ is localized at the left (right) edge.} In order to arrive at the left eigenvectors $t_L \leftrightarrow t_R^{{\color{black}\ast}}$ have to be interchanged.\\
Equation (\ref{gamma1nh}) [(\ref{gamma2nh})] ensures the existence of (\ref{firstev}) [(\ref{secondev})] and thus is a consistency relation. Since \textcolor{black}{$\Gamma_c^a$ and $\Gamma_c^b$} cannot be  $\in[0,1]$ at the same time (apart from the special case $|t_L|=|t_R|$), the amplitude $\psi_n$ on every second site has to vanish and we only find one linearly independent eigenvector. However, the eigenvalue $E=0$ is at least two-fold degenerate, as we have pointed out in Section \ref{Chap_models}, and we thus encounter an EP. 
\\
In Ref. \cite{gong2018topological} it was argued that the Hatano-Nelson model features a topologically non-trivial phase,  in which an  edge state forms for semi-infinite BC. We however will motivate in the following,  that the Hatano-Nelson model does not have edge states for OBC but is gapless instead, which is confirmed by the numerical calculation of the eigenspectrum with exact diagonalization.  Since the Hatano-Nelson model still has one exact zero-energy eigenmode $|\psi_{R[L]}\rangle$ for an odd number of sites \textcolor{black}{(equivalent to Eq. (\ref{eigenmodeR}) [(\ref{eigenmodeL})])}, that explicitly reads as
\begin{align}
&|\psi_R\rangle=\mathcal{N}_R\sum_{n=1}^{(N+1)/2}\left(-\frac{t_R}{t_L}\right)^{n}c^{\dagger}_{2n-1}|0\rangle \label{eigenmodesNHr}\\ &|\psi_L\rangle=\mathcal{N}_L\sum_{n=1}^{(N+1)/2}\left(-\frac{t^{{\color{black}\ast}}_L}{t^{{\color{black}\ast}}_R}\right)^{n}c^{\dagger}_{2n-1}|0\rangle, \label{eigenmodesNH}
\end{align}
the biorthogonal polarization $P$ \cite{kunst2018biorthogonal} (cf. (\ref{eqn:biopol})) can still be constructed with the simplified  projection operator $\Pi_n=c_n^{\dagger}|0\rangle\langle0|c_n$. One already sees that the condition  (\ref{floreGapC}) is always met \textcolor{black}{marking the point where the biorthogonal polarization $P$ jumps between $0$ and $1$}. The  biorthogonal projection of these exact eigenstates $|\psi_L\rangle, |\psi_R\rangle$ is indeed never localized and thus \textcolor{black}{$P=1/2$} (see appendix A). \\
The only exception is the case if at least one of the hoppings $t_L, t_R$ equal zero. At this point, one of the exact eigenstates is the zero vector and the other becomes ill defined. Hence, $P$ is not defined. In fact, here we encounter an (real-space) EP \textcolor{black}{of order $N$ with the $N$-fold degenerate}  eigenenergy $E=0$ (see Figure \ref{FigNelsonHat}a)) and a single eigenvector. Note that the effect of higher order EPs on the failure of the conventional BBC has been discussed \cite{xiong2018does,Martinez2018}.
\\
\textcolor{black}{Since no edge states are found in the Hatano-Nelson model,  the zero-energy eigenmode  in Eq. (\ref{firstev}), [(\ref{secondev})] for a given $t_L, t_R$ always concurs with the bulk gap-closing point if the critical boundary conditions $\Gamma^{a[b]}_c(t_L, t_R)$ are imposed (see Fig. \ref{FigNelsonHat}b) for an example)}. Fig. \ref{FIGgamma}b) thus visualizes equations (\ref{gamma1nh}) and (\ref{gamma2nh}) describing  how the gap-closing points move through parameter space when tuning $\Gamma$ from PBC to OBC.

\paragraph{Non-Hermitian SSH-Model.}
Solving the right eigenvalue problem $H_{SSH}|\psi^R_0\rangle=E|\psi^R_0\rangle$ with the ansatz $|\psi^R_0\rangle=\sum_{n=1}^N\left(\psi^R_{n,A}c^{\dagger}_{n, A
}+
\psi^R_{n,B}c^{\dagger}_{n, B}\right)|0\rangle$  and assuming that the eigenenergy is given by E=0, we arrive at the  following set of equations
\begin{align}
&\psi^R_{n+1, A}=-\frac{t_1-\frac{\gamma}{2}}{t_2}\psi^R_{n, A}, \quad\Gamma \psi^R_{1, A}=-\frac{t_1-\frac{\gamma}{2}}{t_2}\psi^R_{N, A} \label{sysA}\\
&\psi^R_{n-1, B}=-\frac{t_1+\frac{\gamma}{2}}{t_2}\psi^R_{n, B}, \quad\Gamma \psi^R_{N, B}=-\frac{t_1+\frac{\gamma}{2}}{t_2}\psi^R_{1, B}.\label{sysB}
\end{align}
with $n=1,\dots, N-1$ in $(\ref{sysA})$ and  $n=2,\dots, N$ in $(\ref{sysB})$. The system of equations is decoupled with respect to the amplitudes on $A(B)$-sites, such that solving for $\Gamma$ gives two solutions
\begin{align}
\Gamma^A_c=\left(-\frac{t_1- \frac{\gamma}{2}}{t_2}\right)^{N}, \quad \Gamma^B_c=\left(-\frac{t_1+\frac{\gamma}{2}}{t_2}\right)^{N} \label{gammas}
\end{align}
\textcolor{black}{corresponding to the two eigenvectors 
	\begin{align}
	&|\psi^R_{0, A}\rangle=\mathcal{N}_{R}\sum_{n=1}^N\left(-\frac{t_1-\frac{\gamma}{2}}{t_2}\right)^nc^{\dagger}_{n, A}|0\rangle \label{leigenstateAssh}\\
	&|\psi^R_{0, B}\rangle=\mathcal{N}_{R}\sum_{n=1}^N\left(-\frac{t_1+\frac{\gamma}{2}}{t_2}\right)^{N+1-n}c^{\dagger}_{n, B}|0\rangle. \label{leigenstateBssh}
	\end{align}}
The left eigenvalue problem $H^{\dagger}|\psi^L\rangle$ interchanges $t_1-\gamma/2 \leftrightarrow (t_1+\gamma/2)^{{\color{black}{\ast}}}$ in \textcolor{black}{the above equations}.\\
The existence of the eigenvalue $E=0$ is guaranteed by equation (\ref{gammas}): Given a set of parameters $\{t_1, t_2, \gamma\}$ such that $\Gamma^{A(B)}_c \in [0,1]$, we find  the zero-energy eigenstate $|\psi^R_{0, A(B)}\rangle$ exactly at the  boundary condition $\Gamma^{A(B)}_c$.
\\
Similarly to the Hatano-Nelson model, $\Gamma^A_c \not =\Gamma^B_c$  if  $\gamma \not = 0$, forcing the amplitude on every other site to be zero in order to simultaneously satisfy (\ref{sysA}) and (\ref{sysB}). The remaining non-zero amplitudes are  defined apart from a normalizing factor, leading to only one  linearly independent right eigenvector $|\psi^R_{0, A}\rangle$ \textcolor{black}{or $|\psi^R_{0, B}\rangle$} and we encounter an EP. The order of this EP, i.e. the difference between algebraic and geometric multiplicity plus one, is at least two and $E=0$ is at least two-fold degenerate. With exact diagonalization techniques, in can be observed that the EP at $E=0$ in this model is in most cases indeed of order two. One exception can be found at the fine-tuned point $t_2=\gamma/2$ and $t_1=0$. \\
Several remarks are in order. First, we note that the EP found here is  not necessarily the only one but others might occur at different  eigenenergies $E \ne 0$. One known example is the $(N-1)$-fold degenerate bulk-EP at $E=\pm1$ that appears for OBC ($\Gamma=0$)  when $|t_1|=\gamma/2$ \cite{Martinez2018,xiong2018does} that is accompanied by an EP  of order 2 with energy $E=0$, which can be seen from (\ref{gammas}). Second, we stress that the occurrence of the EP depends on the exact number of sites: If it is odd for instance, the eigenenergy $E=0$ is no longer degenerate. 

Rewriting (\ref{gammas}), we arrive at a form that stresses the great sensitivity of the eigenspectrum towards the boundary conditions (compare also to the corresponding Eq.~(\ref{gamma1nh}) for the Hatano-Nelson model)
\begin{align}
\Gamma_c^{A[B]}=e^{N\ln\left(-\frac{t_1-[+]\frac{\gamma}{2}}{t_2}\right)}=e^{-\frac{N}{\xi_{A[B]}}}, \label{GammaExp}
\end{align}
where the argument of the logarithm must be $\leq1$, such that the logarithm is negative.   
\textcolor{black}{$\xi_{A[B]}$ equals the localization length of the zero-energy eigenmode described in Eq. (\ref{leigenstateAssh}) [(\ref{leigenstateBssh})]}. 

The NH SSH model features edge states in the open system \cite{kunst2018biorthogonal,yao2018edge}. In our analysis, we follow the zero-energy mode appearing at the critical value of the boundary conditions  (see Eq. (\ref{gammas},\ref{GammaExp})) through parameter space. Thereby we are able to understand the transformation of the eigenspectrum including the formation of edge states. Since we only follow a single mode, we cannot distinguish between a bulk gap-closing point accompanied by other low-energy modes  (called scenario $(i)$ in the following)  or the formation of isolated edge states in a gapped system (dubbed scenario $(ii)$). To address the issue of telling apart these two scenarios, we additionally compute  the biorthogonal polarization $P$  \cite{kunst2018biorthogonal} (see also Section \ref{Chap_biorth}). Our main observations are summarized in the following.
\\
\textit{Scenario} $(i)$: Whenever the topological phases of the system with PBC (defined by winding numbers) and OBC (defined by polarization $P$) differ from each other (dark green shaded areas in Figure \ref{FigSSHmodel}b)), the bulk gap closes and reopens while tuning $\Gamma$ from $1$ to $0$. Specifically, in this regime either  $\Gamma^{A}_c \in [0,1]$ or  $\Gamma^{B}_c \in [0,1]$  (see Eq. (\ref{gammas},\ref{GammaExp})), but not both. The EP at $E=0$ is then found at the critical value of the boundary conditions which concurs with the bulk gap-closing point.
\\
\textit{Scenario} $(ii)$: The topological properties of the systems with PBC (winding numbers) and OBC (polarization $P$) coincide (light green shaded area in Figure \ref{FigSSHmodel}b)). Then, indeed, the EP at $E=0$ occurring at some critical values of the boundary conditions (in the light-blue area in Fig.~\ref{FigSSHmodel}a) both $\Gamma^{A}_c$ and $\Gamma^{B}_c \in [0,1]$) corresponds to an edge mode forming in the bulk gap. For $\Gamma=0$, the eigenvalue of this edge mode is no longer exactly zero but exponentially small in system size (mini-gap).
\\
\textcolor{black}{At the boundary between the areas described by scenario $(i)$ and scenario $(ii)$ (blue vertical lines in Fig.~\ref{FigSSHmodel}b)), the EP at $E=0$ and $\Gamma_c$ concurs with the gap-closing point, but in contrast to scenario (i) the gap remains closed for $\Gamma<\Gamma_c$. Hence, the bulk gap-closing points for OBC (where the polarization $P$ jumps) are found at these boundaries.  }

\paragraph{General Observations.}
\textcolor{black}{Interestingly,  a connection between the critical boundary conditions $\Gamma_c$ and  the biorthogonal Polarization $P$ can be established for both models, because the zero-energy eigenstates $|\psi^R_{0,A[a]}\rangle$ and $|\psi^R_{0,B[b]}\rangle$  of the NH SSH model  [Hatano-Nelson model] in Eq. (\ref{leigenstateAssh}) and (\ref{leigenstateBssh}) [Eq. (\ref{firstev}) and (\ref{secondev})] resemble the right and left eigenstate  $|\psi_R\rangle$, $|\psi_L\rangle$ in Eq. (\ref{eigenmodeR}), (\ref{eigenmodeL}), which are used for the construction of the polarization $P$. That is, the localization length $\xi_R$ of $|\psi_R\rangle$ coincides with the  localization length $\xi_{A[a]}$ and the localization length $\xi_L$ of $|\psi_L\rangle$ with  $\xi_{B[b]}$. Therefore,  a reformulation of Eq. (\ref{floreGapC}) is given by $\Gamma_c^{A(a)}\cdot \Gamma_c^{B(b)}=1$ that  matches with the condition that the biorthogonal Polarization $P$ needs to jump between 0 and 1. Analogously,  $\Gamma_c^{A(a)}\cdot \Gamma_c^{B(b)}<1$ specifies the parameter regime for which edge states occur for OBC.} The case in which the left and right eigenvectors $|\psi_R\rangle$, $|\psi_L\rangle$  equal each other is reflected by  $\Gamma^A_c=\Gamma^B_c$. One example is given by tuning $\gamma=0$ in the NH SSH model (that is either the Hermitian limit if $t_1,~ t_2 \in \mathbb{R}$ or if $t_1,~ t_2 \in \mathbb{C}$ a version of the model discussed in Ref. \cite{lieu2018topological}), and  one generally can construct two linearly independent eigenvectors except from the cases $\Gamma=1$ and $\Gamma=0$. Then, the  eigenspectra for OBC and PBC coincide apart from edge modes \cite{kunst2018biorthogonal,lieu2018topological}. 

\begin{figure}
	\centering
	\resizebox{0.51\textwidth}{!}{%
		\includegraphics{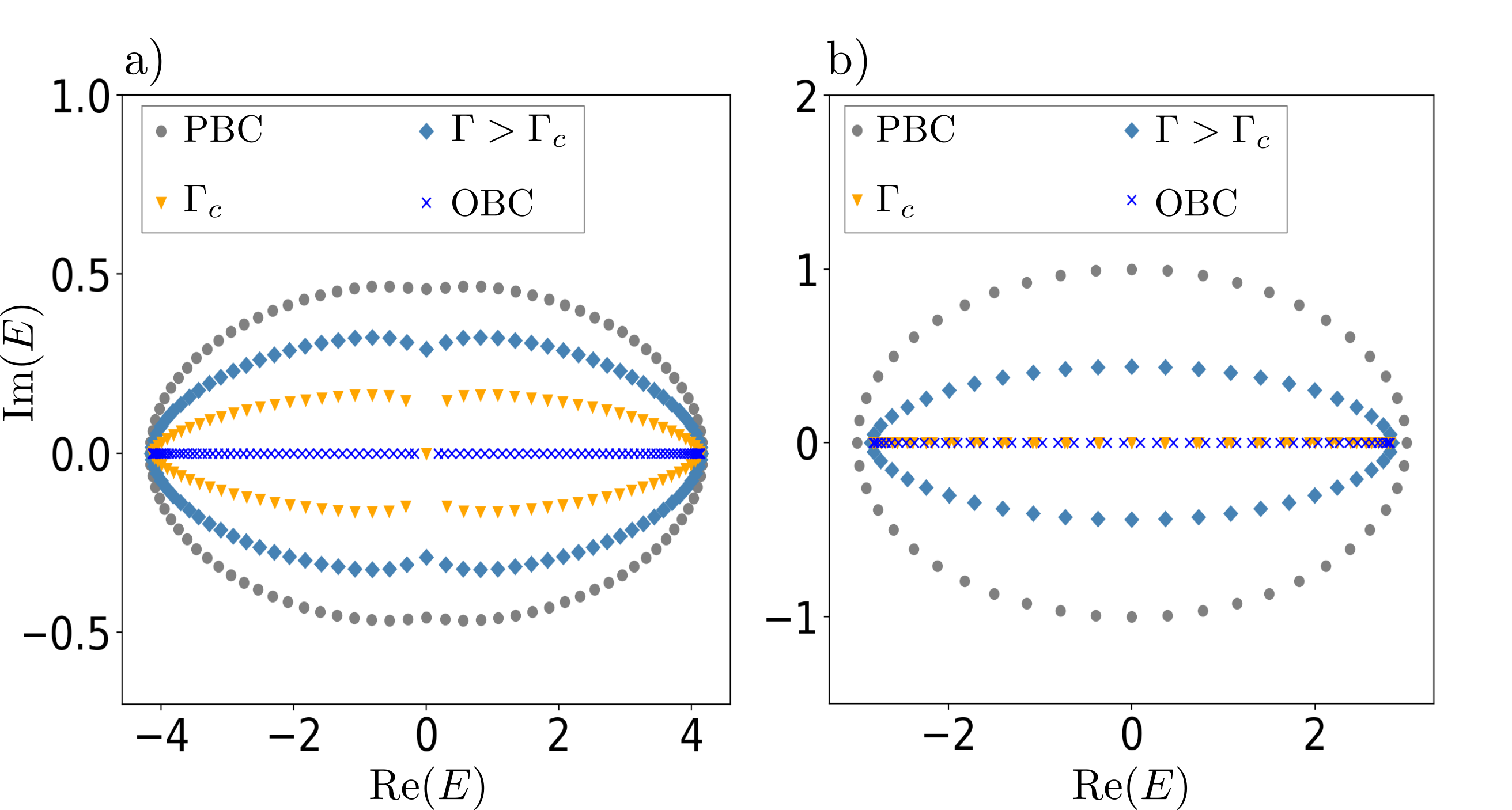}
	}
	\caption{The eigenenergies of the NH SSH model in a) with $N=48$, $t_1=2.2$, $t_2=2$, $\gamma=1$,  and of the Hatano-Nelson model with $N=48$, $t_R=2$, $t_L=1$ in b) are plotted for different boundary conditions in the complex plane. The critical coupling $\Gamma_c$ (in a) given by $ \Gamma^A_c$ in eq. (\ref{gammas}), in b) by $ \Gamma^a_c$ in eq. (\ref{gamma1nh}) terminates the winding around the origin $E=0$.}
	
	\label{FIGwinding}  
	\end{figure}

Quite remarkably, for both the Hatano-Nelson model and the NH SSH model, we find that the eigenenergy spectrum stops winding around the origin $E=0$ in the complex plane precisely at the critical boundary conditions $\Gamma_c$ (see Fig.~\ref{FIGwinding}). That  is, if the eigenenergy winds around the origin for PBC, it does for any $\Gamma>\Gamma_c$. Since we have searched for the boundary conditions that feature an exact zero-energy eigenmode, it is clear that a topological phase transition in the sense of Ref. \cite{gong2018topological} occurs if the base energy (here $E=0$) is crossed as a function of the boundary condition parameter $\Gamma$. Even though we derived the zero-energy eigenmodes for an even number of unit cells, this statement is general and valid for an odd number of cells as well. This finding provides an intuitive explanation for the discrepancy between the topological phase diagram for systems with PBC vs. systems with OBC, namely due to a topological phase transition in which the boundary condition parameter $\Gamma$ plays the role of a control parameter. Thus, when extended to systems with generalized boundary conditions, the spectral winding number displayed in Fig.~\ref{FIGwinding} captures the breakdown of the conventional BBC.  

\section{Stability  of Edge States}\label{Chap_stability}
\begin{figure}
	\resizebox{0.5\textwidth}{!}{%
		\includegraphics{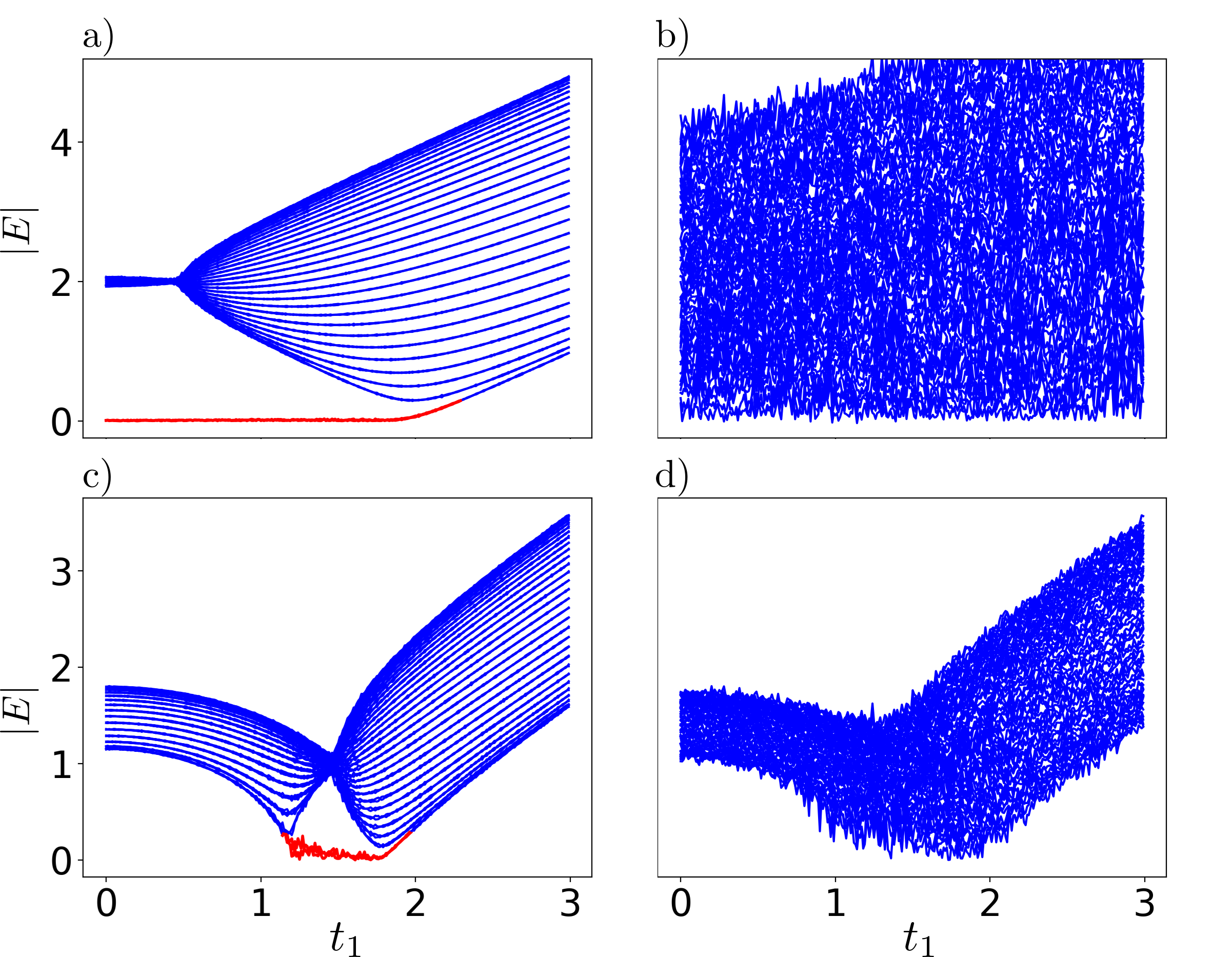}
	}
	\caption{The absolute eigenenergies of the non-Hermitian SSH-model with OBC, $N=30$ and  $t_2=2, \gamma=1$ in a), b)  and  $t_2=1, \gamma=3$ in c), d) are shown in dependency of the hopping $t_1$. A random  complex number $a+i\cdot b$ with $a, b  \in [-1,1]$ scaled with $\exp(-|i-j|^2)$ in a),c) and with  $\exp(-|(i-j)|/\xi_A)$ in b),d), where $t_1=\sqrt{t_2^2+\gamma/2} $ is fixed in $\xi_A$,  is added to the $(i,j)$th and its complex conjugate to the $(j,i)$th matrix element  if it is zero and $i\not \in j$.  While edge modes (shown in red) exist in the first case and thus are robust against these perturbations, they do not persist in the latter case. }\label{FigPertrurbedSSH}     
	\end{figure}

We have seen that the eigenspectrum changes exponentially fast in system size when tuning its boundary conditions. In particular, we have seen for two examples that EPs at $E=0$ start to move in parameter space as soon as the boundary conditions are modified. Since they coincide with the bulk gap-closing point in a wide parameter regime, their position in parameter space as a function of boundary conditions is an indicator for the transition between the qualitative PBC and OBC spectra. Hence, this transition goes with $\propto \exp(-N)$ \cite{kunst2018biorthogonal,xiong2018does,lee2019anatomy} where $N$ is the number of unit cells. This exponential sensitivity raises the natural question whether these apparently fragile spectral properties of systems with OBC are practically observable in realistic systems. On the other hand, an (unwanted) coupling amplitude between the first and the last site in a generic tight-binding model with linear geometry is also strongly suppressed in system size, and we encounter a problem of competing scales that we will take a closer look at in the following.
\\
The dramatic changes in the  eigenspectrum of non-Her\-mi\-tian Toeplitz-matrices (i.e. matrices with constant entries on all diagonals) and operators towards small perturbations have been already discussed in a mathematical context  \cite{reichel1992eigenvalues,gong2018topological}. In particular, it was proposed to rather study so called $\epsilon$-pseudoeigenvalues, that can become proper eigenvalues upon adding a suitable perturbation of norm $\epsilon$. This argument was used in Ref. \cite{gong2018topological} to support the idea of investigating pseudo quasi edge states that stem from imposing half-infinite boundary conditions. As a complementary approach accounting for this spectral instability, Ref.~\cite{herviou2019restoring} proposed to use the more stable singular value spectrum as and alternative to the directly observable eigenspectrum. 
\\
Here, instead we argue that  the considered perturbations should be physically motivated, and are by no means expected to be arbitrarily non-local random matrices as considered in Ref.~\cite{reichel1992eigenvalues}. Instead, realistic unwanted perturbations in a one-dimensional experimental setting with linear chain geometry are for instance given by couplings  between sites with some larger distance that are omitted in a generic tight-binding model. Couplings that would effectively change the boundary conditions are then proportional to the overlap of the Wannier functions centered on the first and the last site. To quantify the generic scaling of this overlap, we may approximate the potential wells in the tight-binding formalism with harmonic potentials, where it is easy to see that the hopping between two  sites $i$ and $j$ scales as $\propto\exp(-|i-j|^2)$ (see  appendix B). Thus, the matrix element coupling the last with the first site is of order $\exp(-N^2)$ which for large enough systems is drastically smaller than the scaling function $\Gamma_c\propto \exp(-\alpha N)$ that describes the transition between the OBC and the PBC  spectrum.
\\
To include this in our analysis of the stability,  we add random perturbations  with the only two constraints that they should be Hermitian and physically motivated. To be precise, we change  all off-diagonal  zero matrix elements of $H$ to $H_{ij}=H^{\ast}_{ji}=(a+ib)\exp(-|i-j|^2)$ where $a,b$ are random numbers $ \in [-1.5, 1.5] $ and thus  are of the same  order as the nearest neighbor hopping amplitudes in our examples in Figure \ref{FigPertrurbedSSH}. We hereby show that the open eigenspectrum as well as the edge states are robust against  disorder that is physically motivated (see Figure \ref{FigPertrurbedSSH}a) and c)). 
\\
To underline the importance of the locality of perturbations, we demonstrate that the edge states are not robust against perturbations that decay $\sim\exp(-|i-j|)$, i.e. comparable to the scaling of $\Gamma_c$. Specifically, adding perturbations that scale as  $\exp(-|i-j|/\xi_A)$, where the factor  $ \xi_A=\left|1/\ln\left(\frac{t_1-\gamma/2}{t_2}\right)\right|$ ensures that especially the perturbation term between the first and the last site is of the same order as the sensitivity of the eigenspectrum. We fix the free parameter $t_1$  of the localization length $\xi_A$ to $t_1=\sqrt{t_2^2+\gamma/2}$  in our example and find, that  boundary modes disappear for such perturbations (see Figure \ref{FigPertrurbedSSH}b) and d)). 

\section{\label{discussion}Discussion}
We have studied the sensitivity of the eigenspectrum  of non-Hermitian tight-binding models towards their boundary conditions. In particular, we derived analytical expressions showing how exact zero-energy modes move through parameter space as a function of the boundary condition parameter $\Gamma$ interpolating between PBC and OBC. In several non-Hermitian models, we find that this eigenmode at $E=0$ represents an EP that can switch from embodying a bulk gap-closing point to being an isolated edge mode. These two cases are found to be distinguished by the biorthogonal polarization $P$ introduced in Ref. \cite{kunst2018biorthogonal}. Corroborating earlier observations, we analytically quantify how at a critical value $\Gamma_c\propto \exp(-\alpha N)$ \cite{kunst2018biorthogonal,xiong2018does,lee2019anatomy} a transition occurs between the qualitative spectral properties found for PBC and the ones for OBC. 

We have further discussed the stability of surface states arising in these models towards physically motivated perturbations. In summary, the spectral instability of non-Hermitian matrices loses its dramatic appearance when imposing physical locality conditions on the considered perturbations. Specifically, in a flat sample geometry, accidental couplings between opposite ends are found to scale as $\propto \exp(-\lambda N^2)$ in generic tight-binding models, and thus naturally decay even faster with system size than the critical coupling  $\Gamma_c$. Moreover, the topological zero-energy edge modes in non-Hermitian systems are found to be robust towards generic local perturbations such as disorder potentials. Hence, despite the sensitivity of non-Hermitian eigenspectra to the choice of boundary conditions, the non-Hermitian BBC as characterized by the biorthogonal polarization $P$ and the concomitant edge modes may be seen as a topologically stable  and experimentally observable phenomenon. 

Finally, we note that some experiments on classical systems such as optical meta-materials, despite a certain similarity to quantum-mechanical tight-binding Hamiltonians, exhibit couplings between sites that do not follow a Gaussian (sometimes not even exponential) decay with distance. Investigating possible modifications to the definition of topological invariants and the stability of the non-Hermitian BBC in such long-ranged settings is an interesting subject of future research.  

\section*{Acknowledgments}
We acknowledge discussions with Emil J. Bergholtz and Flore K. Kunst, as well as financial support from the German Research Foundation (DFG) through the Collaborative Research Centre SFB 1143 and the Cluster of Excellence ct.qmat.

\bibliographystyle{unsrt}

\bibliography{my-bib}
\section*{Appendix A}
\label{appendixP}
To calculate the biorthogonal polarization $P$ for the Hatano-Nelson model, we first find the biorthogonal normalization $\mathcal{N}^{\ast}_L\cdot \mathcal{N}_R$ of the exact eigenstates $|\psi_R\rangle, |\psi_L\rangle$ (see equations (\ref{eigenmodesNHr}), (\ref{eigenmodesNH}))
\begin{align}
&1\equiv \langle\psi_L|\psi_R\rangle=\mathcal{N}^{\ast}_L \mathcal{N}_R\sum_{j}^{\frac{N+1}{2}}\left(\frac{t_Lt_R}{t_Rt_L}\right)^j=\mathcal{N}^{\ast}_L \mathcal{N}_R\left(
\frac{N+1}{2}\right).
\end{align}
Using this normalization condition, we can calculate the polarization $P$ (\ref{eqn:biopol})
\begin{align}
\nonumber&1-\lim\limits_{N\rightarrow\infty}\frac{1}{N}\langle\psi_L|\sum_{j=1}^N\Pi_j|\psi_R\rangle\\
\nonumber&=1-\lim\limits_{N\rightarrow\infty}\frac{1}{N}\langle\psi_L|\sum_{j=1}^N c^{\dagger}_j |0\rangle \langle 0|c_j|\psi_R\rangle\\
\nonumber&=1-\lim\limits_{N\rightarrow\infty}\frac{1}{N}\frac{1}{\frac{N+1}{2}} \sum_{n=1}^{\frac{N+1}{2}}(2n-1)\\
&=1-\lim\limits_{N\rightarrow\infty}\frac{1}{N}\frac{N+1}{2}=\frac{1}{2}.
\end{align}
We thus see, that the biorthogonal polarization $P$ 'jumps' ~betwenn $0$ and $1$, thereby marking the gapless region. 
\section*{Appendix B}\label{appendix}
Under the assumption that the electrons are tightly bound to the atoms, one  assumes within the framework of the tight-binding approximation (compare to e.g. \cite[pp. 56-59]{nolting7}) that the many-body Hamiltonian can be written as a sum
\begin{eqnarray}
H(\mathbf{r})=\sum\limits_{\mathbf{R}_n}H_{\text{atom}}(\mathbf{r}-\mathbf{R}_n)+V_1(\mathbf{r})
\end{eqnarray}
consisting of the Hamiltonians $H_{\text{atom}}(\mathbf{r}-\mathbf{R}_n)$  of the single atoms at positions $\mathbf{R}_n$ and  a correction potential $V_1(\mathbf{r})$. \\
After a Bloch-wave ansatz for the eigenfunctions one arrives at 
\begin{eqnarray}
E_n(\mathbf{k})=E_n +\frac{v_n+\frac{1}{\sqrt{N}}\sum_{j\not=0}\gamma_n^{(j)}e^{i\mathbf{k}\mathbf{R}_j}}{{1+\frac{1}{\sqrt{N}}\sum_{j\not=0}\alpha_n^{(j)}e^{i\mathbf{k}\mathbf{R}_j}}} \label{en}
\end{eqnarray}
with 
\begin{eqnarray}
v_n&=&\int\text{d}^3rV_1(\mathbf{r})|\psi_n(\mathbf{r})|^2\\
\gamma_n^{(j)}&=&\int\text{d}^3r\psi^{\ast}_n(\mathbf{r})V_1(\mathbf{r})\psi_n(\mathbf{r}-\mathbf{R}_j)\\
\alpha_n^{(j)}&=&\int\text{d}^3r\psi^{\ast}_n(\mathbf{r})\psi_n(\mathbf{r}-\mathbf{R}_j)
\end{eqnarray}
where $E_n$ is the $n$th eigenenergy and $\psi_n$ the $n$th eigenfunction of the single atom Hamiltonian. $\gamma_n^{(j)}$ (and $\alpha_n^{(j)}$) translate furthermore into the matrix elements of a tight-binding Hamiltonian in second quantization. Usually, these integrals  become very small and sums in (\ref{en}) are truncated  for small $j$ already. For our purposes, we are interested in the scale describing the decay of these neglected matrix  elements. \\
We approximate the single atoms with  harmonic potentials since  close to the equilibrium point it is a justified assumption for any potential and the eigenfunctions of the quantum oscillator  are well known. Furthermore we assume $V_1$ to be constant so that we effectively only need to study the overlap integral $\alpha_n^{(j)}$. \\
Here, we use the eigenfunctions of the one-dimensional harmonic oscillator,  noting that approximating with a 3D harmonic potential leads to the same asymptotic behavior. The overlap integral between two  sites at positions $\mathbf{R}_i=0$ and  $\mathbf{R}_j=ja$, where $a$ is the lattice spacing, in one spatial dimension is given by
\begin{eqnarray}
\alpha_n^{(j)}=\int\limits_{-\infty}^{+\infty}\text{d}x\psi_n(x)\psi_n(x-j a)
\end{eqnarray}
with eigenfunctions  \cite[p. 312-315]{nolting5_1}
\begin{align}
\psi_n(x)=\frac{1}{\sqrt{2^n n!}}\left(\frac{m\omega}{\pi \hbar}\right)^{1/4}e^{\left(-\frac{m\omega}{\hbar}\frac{x^2}{2}\right)}\cdot H_n\left(\sqrt{\frac{m\omega}{\hbar}}x\right)
\end{align}
and $H_n(z)$ are Hermite polynomials
\begin{eqnarray}
H_n(z)=(-1)^n e^{z^2}\frac{\text{d}^n}{\text{d}z^n}e^{-z^2}. 
\end{eqnarray}
After the typical variable transformation $z=\sqrt{\frac{m\omega}{\pi \hbar}}x$ and shifting the integrand to a symmetric form we find

\begin{align}
\alpha_n^{(j)}=\frac{e^{-\frac{j^2a^2}{4}}}{2^nn!\sqrt{\pi}}\int\limits_{-\infty}^{+\infty}\text{d}ze^{\left(-\frac{z^2}{2}\right)}H_n(z+j a /2)H_n(z-j a/2).
\end{align}
We want to examine the case of large distances $j$ and only consider the first couple of  eigenstates with small $n$. Hence, we are interested in the leading order of the Hermite polynomials $H_n(z)$  in terms of $j a/2$ which  is $2^nz^n$ and thus, the leading term of the integrand will be $2^{2n}(z^2-j^2a^2/4)^n$. Splitting the integrand in summands with different power of $z$, we see that the zeroth power of $z$ is of leading order and explicitly reads
\begin{eqnarray}
\alpha_n^{(j)}\approx\frac{2^{\frac{2n+1}{2}}}{n!}\left(\frac{-j^2 a^2}{4}\right)^ne^{-\frac{j^2a^2}{4}}.
\end{eqnarray}
The matrix element $H_{ij}$ thus scales with $\propto\exp(-|i-j|^2)$. 
\end{document}